\documentclass[aps,pra,twocolumn,showpacs,superscriptaddress,floatfix]{revtex4-1}
\pdfoutput=1
\usepackage{graphicx}
\usepackage{amssymb,amsmath}

{\catcode`\|=\active
\gdef\Braket#1{\left<\mathcode`\|"8000\let|\BraVert {#1}\right>}}
\def\BraVert{\egroup\,\vrule\,\bgroup}

\newcommand{\ket}[1]{\ensuremath{\lvert{#1}\rangle}}
\newcommand{\bra}[1]{\ensuremath{\langle{#1}\rvert}}
\newcommand{\braket}[1]{\ensuremath{\langle{#1}\rangle}}

\usepackage{hyperref}
\hypersetup{colorlinks=true, linkcolor=black, citecolor=black, urlcolor=blue}

\graphicspath{{figures/}{figures/converted/}}

\newcommand{\topp}[1]{\ensuremath{^{({#1})}}}

\DeclareMathOperator{\sinc}{sinc}

\DeclareMathOperator{\indicator}{\mathbf{1}}

\renewcommand{\vec}[1]{\mathbf{{#1}}}
\newcommand{\qop}[1]{\hat{{#1}}}
\newcommand{\qoP}[1]{\ensuremath{\mathcal{#1}}}
\newcommand{\qoT}[1]{\ensuremath{\tilde{\mathcal{#1}}}}

\newcommand{\kvac}{\ensuremath{\ket{\text{vac}}}}
\newcommand{\rhoeff}{\rho_{\text{eff}}}

\newcommand{\cavc}{\text{c}}

\begin{document}
\title{Dynamics of the collective modes of an inhomogeneous spin ensemble in a cavity}

\author{Janus H.~Wesenberg}
\email{janus.wesenberg@nus.edu.sg}
\affiliation{Centre for Quantum Technologies, National University of Singapore, Singapore 117543}
\author{Zoltan Kurucz}
\altaffiliation[Also at ]{Research Institute for Solid State Physics and Optics, H.A.S., H-1525 Budapest, Hungary}
\affiliation{Lundbeck Foundation Theoretical Center for Quantum System Research, Department of Physics and Astronomy, University of Aarhus, DK 8000 Aarhus C, Denmark}
\author{Klaus M{\o}lmer}
\affiliation{Lundbeck Foundation Theoretical Center for Quantum System Research, Department of Physics and Astronomy, University of Aarhus, DK 8000 Aarhus C, Denmark}

\date{\today}

\begin{abstract}
We study the excitation dynamics of an inhomogeneously broadened spin ensemble coupled to a single cavity mode.
The collective excitations of the spin ensemble can be described in terms of generalized spin waves and, in the absence of the cavity, the free evolution of  the spin ensemble can be described as a drift in the wave number without dispersion. In this article we show that the dynamics in the presence of coupling to the cavity mode can be described solely by a modified time evolution of the wave numbers.  In particular, we show that collective excitations with a well-defined wave number pass without dispersion from negative to positive valued wave numbers without populating the zero wave number spin wave mode. The results are relevant for multi-mode collective quantum memories where qubits are encoded in different spin waves.
\end{abstract}

\maketitle

\section{Introduction}

The interaction between a single quantum system and a continuum of independent systems constitutes the standard model of dissipation and decoherence. The Weisskopf--Wigner model thus explains Markovian atomic decay due to coupling to the (oscillator) modes of the quantized radiation field, and the Caldeira--Leggett \cite{caldeira83:quantum} and spin-star \cite{breuer04:non-markovian,schlosshauer08:decoherence} models investigate in detail the  non-Markovian features due to the structure of the reservoir density of states.
Equivalent models have been used to describe the interaction of a two-state fermionic particle with a bosonic continuum \cite{lee54:some}, and the interaction of a laser mode with a gain medium~\cite{scully66:quantum,scully67:quantum,stenholm00:how}.
Common to these works is that the focus is on the reduced dynamics of the central system, that is, on the decoherence and decay of the central system, the properties of the renormalized fermionic particle, or the behavior of the laser mode.

While methods in quantum optics have been developed to address properties of the ensemble of radiation modes, such as the optical spectrum of fluorescence \cite{mollow69:power}, only few detailed studies of the ensemble dynamics exist. Recent proposals to manipulate and protect quantum information stored in collective degrees of freedom of ensembles urge a more precise description of the collective evolution. Systems of interest include the nuclear spins in a host material collectively coupled to a nitrogen-vacancy (NV) center \cite{dutt07:quantum} or a quantum dot \cite{kurucz09:qubit}, trapped atoms \cite{petrosyan08:quantum,petrosyan09:reversible} or polar molecules \cite{rabl07:molecular,tordrup08:holographic} or electron spins \cite{wesenberg09:quantum,imamoglu08:cavity-qed} coupled to a transmission line resonator, and ions in optical cavities~\cite{herskind09:realization}.

\begin{figure}[tb!]
  \centering
  \includegraphics[width=1.0\linewidth]{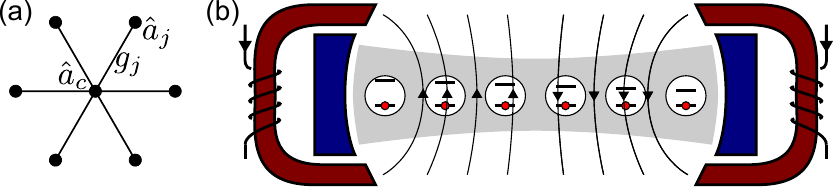}
  \caption{\label{fig:system}(Color online)
    (\textbf{a}) The model studied in this paper describes a set of $N+1$ quantum oscillators with different frequencies interacting in a star topology. We assume that the width of the distribution of the outer oscillator frequencies can be controlled by some external parameter.
    (\textbf{b}) A physical motivation for the model studied is a spin ensemble in a cavity exposed to a controllable linearly varying Zeeman shift provided by an adjustable magnetic field gradient \cite{wesenberg09:quantum}.}
\end{figure}

The inhomogeneities of spins embedded in solids follow typically from properties of the host material, while very well isolated two-level systems may only have different excitation energies if we employ a spatially dependent controllable perturbation to the system. Such a controllable inhomogeneity could be provided by, e.g., a magnetic field gradient. Spin systems with controllable inhomogeneities have been suggested as classical spin echo \cite{anderson55:spin} and photon echo memories \cite{carlson83:storage,kraus06:quantum,afzelius09:multimode}, and it has been proposed that the collective interaction with a cavity could be used to implement a holographic quantum register~\cite{tordrup08:holographic,wesenberg09:quantum}. In the latter, the qubits are stored as collective excitations equivalent to different spin waves, and a Raman process or a linear magnetic field gradient can be applied to deterministically change the wave number between different collective spin wave modes. Such a memory can be read out by reversing the inhomogeneity via applying an inverse Raman process or magnetic field gradient across the ensemble or, as pointed out and demonstrated in Ref.~\cite{wu09:storage}, by employing spin echo techniques to effectively invert the inhomogeneity.

In this paper we consider the collective spin dynamics in an inhomogeneously broadened system of $N$ independent spins $\qop{\vec{S}}_j$ coupled to a single cavity mode as illustrated in Fig.~\ref{fig:system}.
In such a system a particular collective mode of the spin ensemble will experience an enhanced coupling  to the cavity mode due to constructive interference of the single spin coupling terms, and we will refer to this mode as the superradiant spin mode. The coupling to the cavity mode results in an energy gap between the superradiant mode and the other, subradiant, collective modes of the spin system. A sufficiently large energy gap protects the superradiant mode from the dephasing caused by the inhomogeneous precession of the different spin ensemble members, an effect proposed in \cite{kurucz09:qubit} for a nuclear spin ensemble interacting with a single electron spin.
Consider an excitation created in the superradiant mode and left to evolve freely, i.e., without coupling to the cavity field, for a time $T$ much larger than the inverse inhomogeneous width of the spin ensemble. If the cavity coupling is enabled at this time, the excitation will not interact with the cavity because of the destructive interference of the individual spin components.
Reversing the sign of the inhomogeneity or applying a $\pi$-pulse to the entire ensemble will, however, cause the system to refocus into the superradiant mode after another period of length $T$ (spin echo). Note the significance of the cavity coupling during this process: If the cavity mode is not coupled to the spins, the excitation may refocus perfectly into the superradiant mode where it would couple strongly to the cavity mode. If, however, the cavity coupling is turned on during the refocussing period, the energy gap due to the coupling prevents the refocusing into the superradiant mode. Our main goal in the following is to understand the interplay between the  spin-cavity gapping interaction and the inhomogeneity under such conditions.

The paper is organized as follows. In Sec.~\ref{sec:physical-model}, we present the Hamiltonian describing our physical system and introduce the bare- and dressed-time states which are related by an analytical  expression derived in the Appendix. In Sec.~\ref{sec:shear-approximation} we then consider the propagation of a spin wave packet  and introduce a `shear' approximation to provide a concise quantitative explanation of the numerically observed effects. Section \ref{sec:conclusion} concludes the paper.

\section{Physical Model}
\label{sec:physical-model}

A large ensemble of spins prepared with high probability in the spin down eigenstates is conveniently described by the Holstein--Primakoff approximation which replaces the spin operators by bosonic oscillator operators, $\qop{S}_+\topp{j} \approx \qop{a}^\dagger_j$, $\qop{S}_z\topp{j}=\qop{a}_j^\dagger\qop{a}_j - 1/2$, where $\qop{a}^\dagger_j$ is the creation operator for a quantum oscillator representing the $j$-th spin \cite{holstein40:field,kurucz09:qubit} and we set $\hbar=1$. This representation clearly does not reflect the saturation by a single excitation of each individual spin, but with a total number of excitations much smaller than the number of spins, the most significant state components will have at most a single excitation in each oscillator. In this limit, the Holstein--Primakoff approximation is a valid and convenient description of the spin ensemble as a collection of oscillators.

We will assume that the spins are strongly coupled to a single mode of the cavity, and a cavity photon makes multiple round trips before it gets absorbed by the ensemble, so that we can neglect propagation effects in the medium. We further assume that the excitation energies of the spins are much larger than the strengths $g_j$ of the couplings to the cavity mode $\qop{a}_{\cavc}$, so that in the rotating wave approximation the interaction picture Hamiltonian is $\qoP{H}=\qoP{H}_0+\qoP{V}$ where $\qoP{H}_0$ and $\qoP{V}$ describe the free evolution and interaction,
\begin{subequations}
  \label{eq:model}
\begin{gather}
  \qoP{H}_0 \equiv \omega_{\cavc}  \qop{a}_{\cavc}^\dagger \qop{a}_{\cavc}   + \sum_{j=1}^N \omega_j  \qop{a}_j^\dagger \qop{a}_j,\\
  \qoP{V} \equiv
  \sum_{j=1}^N g_j \qop{a}_j^\dagger\qop{a}_{\cavc} +\text{H.c.}
  = \Omega  (\qop{a}_{\cavc} \qop{b}^\dagger +\text{H.c.}).
\end{gather}
\end{subequations}
Here the spin excitation energies  $\omega_j$ and the cavity mode frequency $\omega_{\cavc}$ are defined relative to a common fixed frequency, and $\Omega \equiv (\sum_{j=1}^N \lvert g_j\rvert^2)^{1/2}$ is the collectively enhanced effective coupling strength between the cavity and the superradiant mode described by the creation operator \cite{wesenberg09:quantum}
\begin{equation}
  \label{eq:bdaggerdef}
  \qop{b}^\dagger=\sum_{j=1}^N \alpha_j \qop{a}_j^\dagger
  \text{, where } \alpha_j=\frac{g_j}{\Omega}.
\end{equation}
The system Hamiltonian $\qoP{H}$ is purely quadratic in the bosonic operators and does not describe any interaction between excitations. The dynamics of the non-interacting excitations is therefore completely described by the evolution of the one-excitation subspace, and to simplify the notation in what follows we will consider the evolution of states in that subspace.

\subsection{Bare-time states}

When the cavity is tuned far out of resonance and the spins precess freely in the presence of inhomogeneity, the population in the superradiant state, defined as the singly excited state of the collective spin wave mode, $\ket{0}\equiv \qop{b}^\dagger \ket{\text{vac}}$ gradually becomes subradiant. We will refer to the resulting distinguished subradiant states as bare-time states $\ket\tau$ and label them with the time $\tau$ it takes for the system to reach them under this free evolution starting from the superradiant state,
\begin{equation}
  \label{eq:taudef}
  \ket{\tau} \equiv \qoP{U}\topp{0}(\tau) \qop{b}^\dagger \kvac
  = \sum_{j=1}^N \alpha_j e^{-i \omega_j \tau} \qop{a}_j^\dagger \ket{\text{vac}},
\end{equation}
where $\qoP{U}\topp{0}(\tau) \equiv e^{-i\qoP{H}_0\tau}$ is the free evolution operator. The bare-time states play an important role when the ensemble is used as a multi-mode quantum memory, and we note that if the spins are uniformly distributed along the $z$ axis and $\omega_j =-\kappa  z_j$, the bare-time state $\ket{\tau}$ is the first excited state of the spin-wave with wave number $k=\kappa\tau$ \cite{bloch30:ferromagnetismus}.
In a finite size system, spin wave states, and hence the bare-time states, form an overcomplete basis in the set of singly excited spin states which are accessible from the superradiant state and
\begin{equation}
\label{eq:non-orthogonal-basis}
\braket{\tau_2|\tau_1}=
\sum_{j=1}^N \lvert\alpha_j\rvert^2 e^{-i\omega_j(\tau_1-\tau_2)}
= \int e^{-i \omega (\tau_1-\tau_2)} \rhoeff(\omega) d\omega,
\end{equation}
is proportional to the inverse Fourier transform of the effective spin density defined as the density of spin states weighted by the normalized coupling strength,
\begin{equation}
  \label{eq:rhodef}
  \rhoeff(\omega)\equiv\sum_{j=1}^N \lvert \alpha_j \rvert^2 \delta(\omega-\omega_j).
\end{equation}
In the following we shall be primarily interested in the limit of very many spins, and hence we shall indiscriminately treat the spin excitation energies as an integration variable and a summation index. We will assume that the distribution of energies is such that the overlap $\braket{\tau_2|\tau_1}$, which describes how a collective excitation dephases due to the inhomogeneity, decays on a time scale $T$ on the order of the inverse width of the ensemble.
This in particular implies that  for $|\tau| \gg T$ the bare-time state $\ket\tau$ is orthogonal to the superradiant state $\ket0$, the effect of the cavity on $\ket\tau$ is insignificant, and the time evolution of $\ket\tau$ is the same under the free Hamiltonian $\qoP{H}_0$ as under the full Hamiltonian $\qoP{H}_0+\qoP{V}$.

In the absence of coupling to the cavity, the time evolution of the bare-time states follows from the definition,
\begin{eqnarray}
  \label{eq:disp}
  \qoP{U}\topp{0}(t) \ket{\tau} = \ket{ t+\tau},
\end{eqnarray}
so the free evolution of the inhomogeneously broadened ensemble in the bare-time representation amounts to a displacement of the bare-time argument equivalent to a linear change of the collective spin wave number $k \mapsto k+\kappa t$.

\begin{figure}[tb!]
  \centering
  \includegraphics[width=\linewidth]{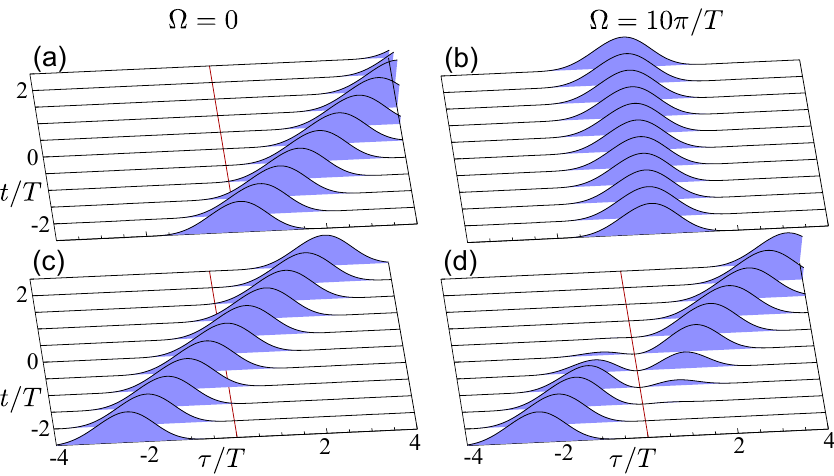}
  \caption{\label{fig:evolution}(Color online)
Time evolution of a collective spin excitation expanded in the bare-time (wave number) basis. The  figures show the overlap $\lvert\braket{\tau|\Psi(t)}\rvert^2$ of the time evolving spin state with the bare-time basis states. Figures (a) and (c) show that the evolution  under the free Hamiltonian $\qoP{H}_0$ ($\Omega=0$) is a simple translation with time. Under the full Hamiltonian $\qoP{H}$ ($\Omega=10\pi/T$), the strongly coupled state $\ket{\tau=0}$ is frozen, (b), while a bare-time state with initial condition $\ket{\Psi(t\to-\infty)} \equiv \ket{\tau=t}$ passes the $\tau=0$ region without ever populating the strongly coupled $\ket{\tau=0}$ state, (d). Snapshots of the expansion of the states in (c) and (d)  are shown in Fig.~\ref{fig:halfwaveprop}.
}
\end{figure}

\subsection{Dressed-time states}
\label{sec:dressed-time-states}

To consider the ensemble dynamics under the full Hamiltonian, we introduce the \emph{dressed-time states} $\ket{\Phi(\tau)}$ as the limit
\begin{equation}
  \label{eq:dresseddef}
  \ket{\Phi(\tau)} \equiv \lim_{t \to \infty} \qoP{U}(t) \qoP{U}\topp{0}(-t) \ket{\tau},
\end{equation}
where $\qoP{U}(t)=e^{-i(\qoP{H}_0+\qoP{V})t}$.
In analogy to the time translation property of the bare-time states as described by Eq.~(\ref{eq:disp}) we have for the dressed-time states
\begin{equation}
  \label{eq:dresseddisp}
  \qoP{U}(t) \ket{\Phi(\tau)} = \ket{\Phi(t+\tau)},
\end{equation}
so that in particular, once we have established the form of $\ket{\Phi(0)}$, we can redefine the dressed-time states as $\ket{\Phi(\tau)} = \qoP{U}(\tau) \ket{\Phi(0)}$.

As the cavity field does not couple to bare-time states with large negative $\tau$, the bare- and dressed-time states coincide in the limit $\tau \to -\infty$. In general, however, the dressed-time states are cavity polaritons with both spin and cavity components. As detailed in the Appendix, we can describe the coupling in terms of twice the complex phase of the inverse cavity mode susceptibility which can be computed as
\begin{equation}
  \label{eq:phidef}
  \phi(\omega) \equiv 2 \arg\left(\left[\omega-\omega_\cavc - \Omega^2 \delta_\cavc(\omega)\right] + i\ \Omega^2 \frac{\gamma_\cavc(\omega)}{2} \right),
\end{equation}
where we have introduced $\gamma_\cavc(\omega) = 2 \pi \rhoeff(\omega)$ and
\begin{equation}
  \label{eq:deltacdef}
  \delta_\cavc(\omega) = \mathcal{P} \int \frac{\rhoeff(\omega')}{\omega-\omega'} d\omega',
\end{equation}
with $\mathcal{P}$ denoting the principal part of the integral.
In general $\phi$ depends on the cavity frequency and coupling strength as well as the effective spin density, but for $\omega$ where $\Omega^2 \gg \lvert (\omega-\omega_\cavc)/(\delta_\cavc(\omega)+i \gamma_\cavc(\omega))\rvert$ we see that $\phi$ is given by
\begin{equation}
  \label{eq:phistrong}
  \phi_\infty(\omega)\equiv-2 \arctan\left(\frac{\gamma_\cavc(\omega)}{2 \delta_\cavc(\omega)}\right)
\end{equation}
independent of these parameters, and we will refer to the system as being in the `strong coupling limit' if this condition is met for all $\omega$ of interest. If the cavity is tuned far outside the support of $\rhoeff$ so that $\Delta_\cavc \approx \omega_c-\omega$, and $\lvert\delta_\cavc(\omega)+i \gamma_\cavc(\omega) \rvert$ is everywhere on the order of $2\pi/T$,  we reach the strong coupling limit when $\Omega^2 \gg \lvert\Delta_c\rvert 2\pi/T$.

\begin{figure}[tb!]
  \centering
  \includegraphics[width=.85\linewidth]{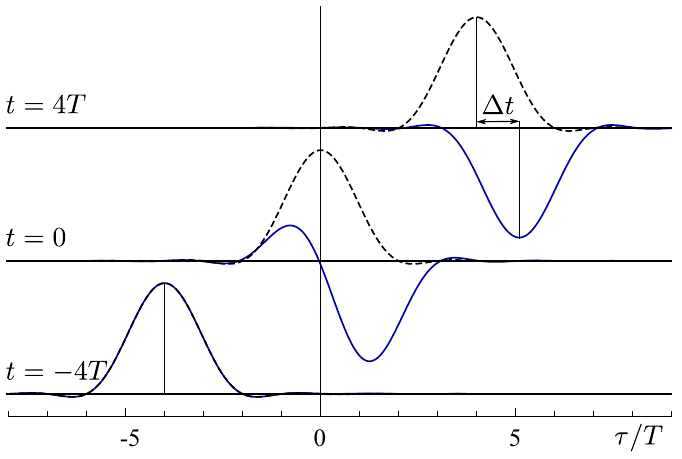}
  \caption{\label{fig:halfwaveprop}(Color online)
Evolution of a wave packet in the bare-time (wave number) basis in the presence and absence of a strongly coupled cavity. The initial state (bottom figure) is the negative bare-time state $\ket{\tau=-4T}$ at $t=-4T$, and the figures show the overlaps $\braket{\tau|\Psi(t)}$ (solid) and $\braket{\tau|\Psi_0(t)}$ (dashed) at three different times during the evolution. With no cavity coupling, the spin wave packet evolution is a simple translation, while, the cavity coupling freezes the superradiant state, and unitarity forces the evolving wave packet to maintain a vanishing overlap $\braket{0|\Psi(t)}=0$ during the evolution. The net effect on the final state (upper figure) is a phase shift $\pi$ and an additional translation of the wave packet by $\Delta t \sim T$.
}
\end{figure}

In terms of $\phi$, we find in the Appendix that the Fourier transform of the dressed-time states can be expanded as
\begin{multline}
  \label{eq:fulldressed}
  \ket{\tilde{\Phi}(\omega)}
   = \frac{i}{\Omega}\left(e^{-i\phi(\omega)}-1\right)\ket{\cavc}\\
  +\int_{-\infty}^0  e^{i \omega \tau} \ket{\tau} d\tau
  + e^{-i \phi(\omega)} \int_0^{\infty}  e^{i \omega \tau} \ket{\tau} d\tau.
\end{multline}
This expansion is readily evaluated for any choice of the spin couplings and density of states, and it can then be transformed back to provide the time dependent state of the system. Fig. \ref{fig:evolution} compares the time evolution of states in the absence and presence of the cavity coupling. The states are expanded on the bare-time (wave number) basis, and we see that the free evolution is a simple translation with respect to the bare-time argument. The cavity coupling, however, freezes the superradiant $\ket{\tau=0}$ state and it significantly modifies the shape of a wave packet during propagation from negative to positive $\tau$, while asymptotically restoring its shape but causing an additional translation of the bare-time (wave number) argument. In Fig. \ref{fig:halfwaveprop}, we show snapshots comparing the coupled and uncoupled evolution of wave packets starting at negative bare-time arguments, and we show that in addition to the extra translational shear, the interaction provides a change of sign of the wave packet. These numerical results were obtained for an inhomogeneously broadened ensemble with $\rhoeff(\omega) = \tfrac{T}{\pi} \cos^2(\omega T/2) \indicator_{\left[-\pi,\pi\right]}(\omega T)$, see  Fig.~\ref{fig:systemgeometry}b. In the subsequent section, we will give an explanation of these observed phenomena and an analytical approximation to the shear and phase shift in terms of the physical parameters of the model, and we will illustrate results for other inhomogeneity models.

\section{\label{sec:shear-approximation}Wave packet propagation}

To explain the results in Figs.~\ref{fig:evolution} and \ref{fig:halfwaveprop}, let us consider a spin wave packet with the asymptotic initial condition
\begin{gather}
  \label{eq:initcond}
 \ket{\Psi(t \to -\infty)} = \int \psi(t-\tau) \ket\tau \, d\tau = \int \psi(t-\tau) \ket{\Phi(\tau)} \, d\tau.
\end{gather}
When there is no coupling to the cavity field, e.g., because the cavity is far detuned from the spin resonance, the spins evolve under the influence of only the free Hamiltonian, $\qoP{H}_0$,
\begin{equation}
  \label{eq:bareevolution}
  \ket{\Psi_0(t)}=
  \int \psi\left(t-\tau\right) \ket{\tau} d\tau,
\end{equation}
as illustrated in Figs.~\ref{fig:evolution}a and~c. Similarly, the evolution under the influence of the full Hamiltonian, $\qoP{H}=\qoP{H}_0+\qoP{V}$,  can be expressed in terms of the dressed-time states as
\begin{align}
  \label{eq:fullevolution}
  \ket{\Psi(t)}
  = \int \psi(t-\tau) \ket{\Phi(\tau)} d\tau.
\end{align}

When the cavity is strongly coupled to the ensemble, the cavity state $\ket{\cavc}$ and the superradiant state $\ket{0}$ practically decouple from all other collective spin states. Therefore the time evolving wave packet (\ref{eq:fullevolution}) is prevented from refocusing into the superradiant state. We observe numerically (see Fig.~\ref{fig:evolution}) that the wave packet smoothly evades the superradiant mode and makes a transition from a superposition of negative to positive bare-time states. To a good approximation the wave packet asymptotically regains its original shape up to a phase shift and a displacement of the expansion of the state on the bare-time basis states as illustrated in Fig.~\ref{fig:halfwaveprop}.

To explain the observed shear effect and show how the displacement and the phase shift is related to the physical parameters of the system,  we will expand the spin component of the wave packet as $\int \chi(\tau,t) \ket{\tau} d\tau$. Using Eq.~\eqref{eq:fulldressed} we find the expansion coefficient
\begin{equation}
  \label{eq:shearexact}
  \chi(\tau,t) =
  \begin{cases}
    \psi(t-\tau) &\text{for $\tau<0$,}\\
    \frac{1}{2\pi}
    \int e^{-i [(t-\tau) \omega+\phi(\omega)]} \tilde{\psi}(\omega)
    d\omega
    &\text{for $\tau>0$.}
  \end{cases}
\end{equation}
This is an exact description of the evolution of a wave packet in the spin ensemble, valid for an arbitrary time~$t$ and arbitrary wave packet envelope $\psi(t)$. In what follows, we assume that $\tilde\psi(\omega)$ has significant value only in a vicinity of $\omega_0$ within the support of $\rhoeff(\omega)$, and that the linear approximation
\begin{equation}
  \label{eq:philin}
  \phi(\omega) \approx \phi(\omega_0) + (\omega- \omega_0)\phi'(\omega_0)
  =\phi_0+ \omega \Delta t
\end{equation}
is valid in this vicinity. In this case, the integral over frequencies in Eq.~\eqref{eq:shearexact} can be approximately evaluated,
\begin{equation}
  \label{eq:shearapprox}
  \chi(\tau,t) \approx
  \begin{cases}
    \psi(t-\tau) &\text{for $\tau<0$,}\\
    e^{-i \phi_0} \psi(t+\Delta t-\tau)
    &\text{for $\tau>0$.}
  \end{cases}
\end{equation}
One readily sees that asymptotically, the wave packet is shifted by $\Delta t$ and acquires a phase $\phi_0$, by the interaction with the cavity mode, and through Eqs.~\eqref{eq:phidef} and \eqref{eq:philin}, these quantities are explicitly given by the physical parameters of the model.

\begin{figure*}
  \centering
  \includegraphics[width=.9\linewidth]{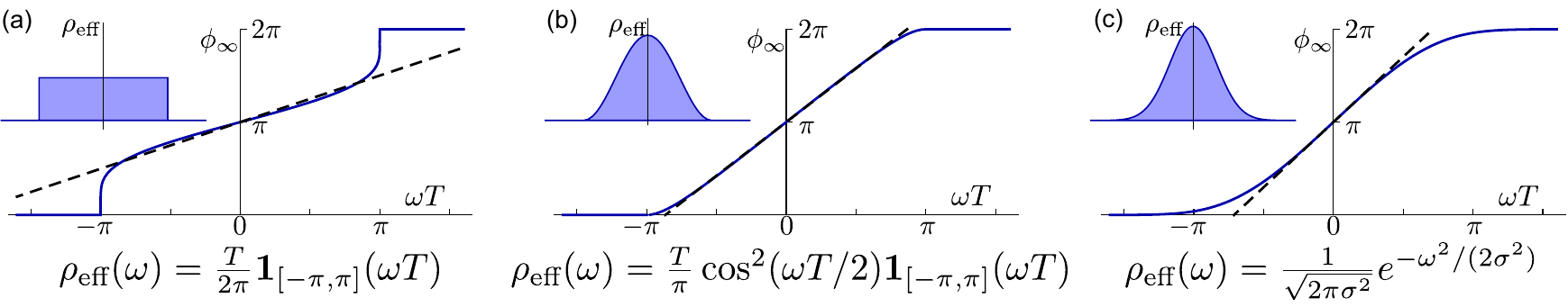}
  \caption{ \label{fig:systemgeometry} (Color online)
Examples of the phase $\phi$ in the strong coupling limit for three different system geometries discussed in the text. For each part (a)-(c), the effective spin density $\rhoeff$ is illustrated by the inset and the plot shows the value of $\phi_\infty(\omega)$ as given by Eq.~(\ref{eq:phistrong}) (solid lines), together with the linear approximation (\ref{eq:philin}) corresponding  to the shear observed for fully delocalized wave packets [$\psi(t)=\delta(t)$] in the system as discussed in Sec. \ref{sec:shear-deloc-wavep} and Fig.~\ref{fig:systemtheory}.
}
\end{figure*}

In the following we will investigate the validity of the shear approximation (\ref{eq:shearapprox}) for three different example systems parametrized by their effective spin density $\rhoeff$:
(a) a uniformly broadened ensemble where the coupling strength of the individual spins is not correlated with the excitation energy of the spins, corresponding to  $\rhoeff (\omega) = \tfrac{T}{2\pi} \indicator_{\left[-\pi,\pi\right]}(\omega T)$.
(b) a uniform spatial distribution of spins with constant magnetic field gradient along a straight halfwave stripline resonator, corresponding to a trigonometric  dependence on frequency, $\rhoeff(\omega) = \tfrac{T}{\pi} \cos^2(\omega T/2) \indicator_{\left[-\pi,\pi\right]}(\omega T)$.
(c) a spin ensemble with a Gaussian density of states, typical for a nuclear spin ensemble in a semiconductor quantum well, where the inhomogeneous broadening is due to the hyperfine interaction with a spin-polarized electron in the well (Knight shift). The frequency variance $\sigma^2=(\pi^2-6)/3 T^2$ is chosen to match the variance of $\rhoeff(\omega)$ for case $b$, the halfwave resonator.
The properties of these three systems are summarized in Fig.~\ref{fig:systemgeometry}.

\begin{figure}
  \centering
  \includegraphics[width=.85\linewidth]{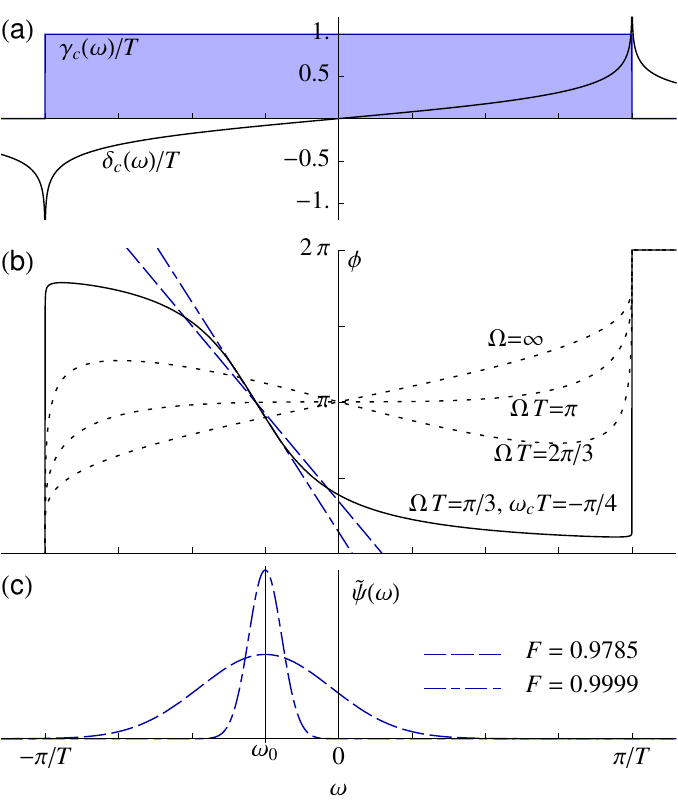}
  \caption{\label{fig:finiteshear}(Color online)
    The shear approximation for a localized wave packet.
    (a) The value of $\gamma_\cavc$ (filled) and $\delta_\cavc$ (solid) as given by Eq.~(\ref{eq:deltacdef}) for the spin system with uniform distribution of $\omega$, $\rhoeff(\omega) = \tfrac{T}/{2\pi} \indicator_{[-\pi,\pi]}(\omega T) $, as also considered in Fig.~\ref{fig:systemgeometry}a.
    (b) The phase $\phi(\omega)$ given by Eq.~(\ref{eq:phidef}) at $\omega_\cavc=0$ and $\Omega=\{2\pi/3, \pi, \infty\}/T$ (dotted) and at $\omega_\cavc=-\pi/4 T$, $\Omega=\pi/3 T$ (solid). Dashed lines are linear approximations of $\phi(\omega)$ according to Eq.~(\ref{eq:philin}) in two different frequency domains around $\omega_0$, where the wave packets illustrated in part (c) of the figure have significant value. The slope of the lines is given by the $\Delta t$ that maximizes the fidelity (\ref{eq:shearfid}).
     (c) The frequency domain wave packets $\tilde{\psi}(\omega)$, and the shear fidelity $F$ for these wave packets.
  }
\end{figure}

\subsection{Shear for localized wave packets}
\label{sec:shear-local-wavep}

From the exact expression (\ref{eq:shearexact}), we expect that the shear approximation (\ref{eq:shearapprox}) is valid if the frequency argument of $\tilde\psi(\omega)$ attains values within a narrow interval, corresponding to states expanded on a wide range of bare-time states. To verify this expectation, we introduce the shear fidelity $F$ as the asymptotic overlap of the exact asymptotic state $\ket{\Psi(t)}$ with the bare-time states $\ket{t+\Delta t}$ maximized over the shift $\Delta t$,
\begin{gather}
  \label{eq:shearfid}
  F \equiv \max_{\Delta t} \lim_{t\to\infty} \lvert \braket{t+\Delta t|\Psi(t)} \rvert.
\end{gather}
As shown in Fig.~\ref{fig:finiteshear}, the size of the optimum shear indeed approaches $\phi'$ as we consider wave packets which are well localized in frequency while at the same time the fidelity is seen to approach unity as the linear approximation to $\phi$ becomes more accurate over the support of the wave function, exactly as we would expect.
The configuration in which the shear approximation is studied in Fig.~\ref{fig:finiteshear} is well outside the strong coupling limit as most clearly seen from the fact that the form of $\phi$ is very different from the strong coupling value ($\Omega = \infty$). This illustrates that high fidelity shear effects can be observed for any coupling strength. Also, the figure indicates that varying the cavity frequency and/or the coupling strength allows very different shear effects, even for a constant system geometry as described by $\rho_{\text{eff}}$.

\begin{figure}
  \centering
  \includegraphics[width=0.85\linewidth]{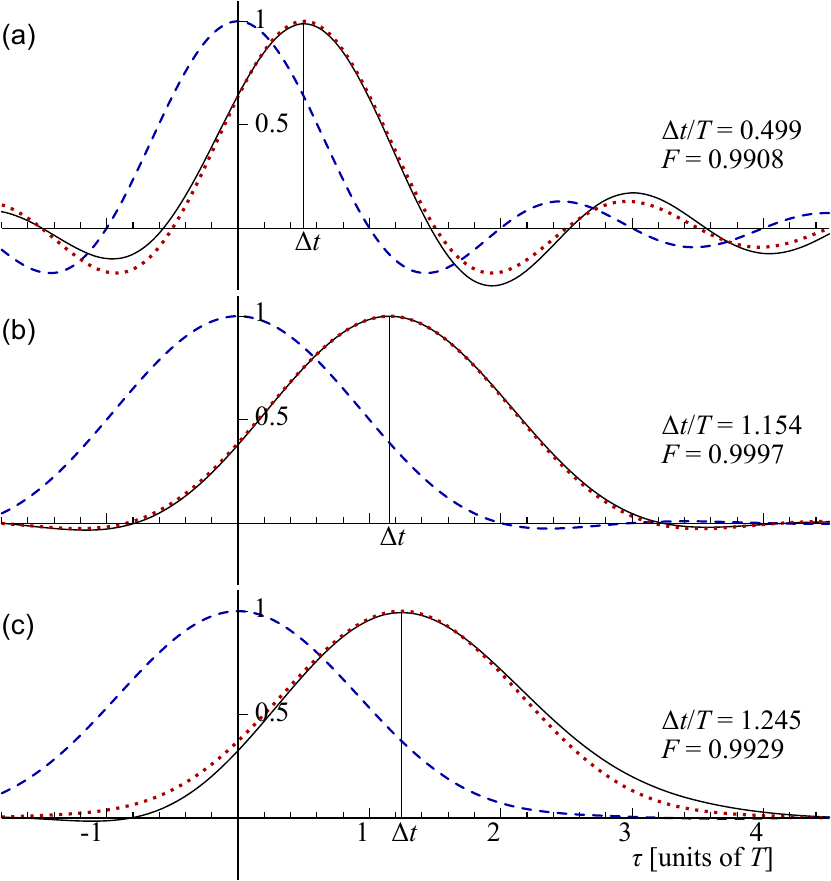}
  \caption{ \label{fig:systemtheory} (Color online)
Shear fidelity for delocalized wave packets. Plot shows the overlap of $\ket{\tau+t}$ with (i) the unshifted state $\ket{t}$ (dashed), (ii) the state $\ket{t+\Delta t}$ corresponding to the shear approximation (dotted), and (iii) the actual asymptotic state compensated for $\phi_0$,  $e^{i \phi_0} \ket{\Psi(t)}$ (solid), for the same three spin systems (a)--(c) as described in Fig.~\ref{fig:systemgeometry} in the limit of $t \to \infty$. For all three systems the wave packet traverses the $\tau\approx 0$ region with little dispersion as discussed in Sec.~\ref{sec:shear-deloc-wavep} and a time shift $\Delta t$ in good agreement with the results of Sec.~\ref{sec:shear-approximation} as illustrated by Fig.~\ref{fig:systemgeometry}.
}
\end{figure}

\subsection{\label{sec:shear-deloc-wavep}Shear for delocalized wave packets}

We have verified above that the shear approximation is indeed valid when $\tilde\psi(\omega)$ only has support over a narrow frequency range where a linear approximation of the phase $\phi$ is reasonable. The examples shown in Figs.~\ref{fig:evolution} and~\ref{fig:halfwaveprop}, however, are for states with well defined $\tau$, [$\psi(t)=\delta(t)$] so that $\tilde{\psi}(\omega)$ is constant and extends to all values of $\omega$, invalidating any linear approximation of $\phi$ as illustrated by Fig.~\ref{fig:systemgeometry}.
In this section our aim is to qualitatively investigate why for some systems we nevertheless observe a relatively high fidelity shear effect as suggested by Figs.~\ref{fig:evolution} and \ref{fig:halfwaveprop} and illustrated quantitatively in Fig.~\ref{fig:systemtheory}.

One way to understand the shear effect for delocalized wave packets is to note that the expansion in Eq.~\eqref{eq:initcond} is not unique: Performing the convolution in frequency space, we can write $\braket{\tau'|\Psi_0(t)}$ as $\int e^{- i \omega (t-\tau')} \tilde\psi(\omega) \rhoeff(\omega) d\omega$, making it clear that only the values of $\tilde\psi(\omega)$ where $\rhoeff(\omega)$ is nonzero are significant.
Therefore, we have the freedom to consider only such initial wave packets for which the support of $\tilde\psi(\omega)$ is within the support of $\rhoeff(\omega)$.
For example, if the support of $\rhoeff(\omega)$ is the interval $[-\pi/T,\pi/T]$ (Figs.~\ref{fig:systemgeometry}a and~b), then instead of $\tilde\psi(\omega) = 1$ corresponding to $\psi(t) = \delta(t)$, we may think of the indicator function $\tilde\psi'(\omega) = \indicator_{[-\pi/T,\pi/T]}(\omega)$, whose inverse Fourier transform is $\psi'(t)=\sinc(\pi t/T)/T$.
This in particular implies that the linear approximation (\ref{eq:philin}) need only be valid for values where $\rhoeff(\omega)$ is nonzero, so that consulting Fig.~\ref{fig:systemgeometry} we would expect a high shear fidelity for case b in the strong coupling limit, as is indeed observed in Fig.~\ref{fig:systemtheory}b. Not surprisingly, Fig.~\ref{fig:systemtheory}c shows that the shear fidelity for case c, where $\rhoeff$ is qualitatively very similar to that of case b, is also high.

\section{Conclusion}
\label{sec:conclusion}
To summarize, we have established a convenient description of the evolution of an inhomogeneously broadened spin ensemble coupled to a central spin or a cavity mode. The dispersion in eigenfrequencies of the individual spins leads to propagation, which is effectively described as a translation in a bare-time basis, equivalent to a spin wave number basis, and for a range of parameters, the joint effect of the inhomogeneity and the coupling of the cavity to a single superradiant mode can be understood in the same basis:  The superradiant mode is gapped and hence becomes stationary (or a dressed state pair of the superradiant spin mode and the cavity mode form coupled, stationary eigenmodes of the system) even in the presence of inhomogeneities. Rather than diagonalizing the problem and finding all eigenmodes of the system, we observe that the bare-time basis states with sufficiently large positive and negative time argument evolve by a simple translation without dispersion in the bare-time basis argument. When the evolution causes this argument to approach the value $\tau=0$ of the superradiant mode, a transient nodal structure forms until the system again becomes a simple bare-time state, but with a changed phase and a shear, that we can compute from the physical parameters of the model.
Since it is a linear effect for highly polarized spin ensembles, the shear occurs not just for single excitations, as studied in this manuscript, but for any quantum or classical excitation of the initial state, provided that the ensemble remains strongly polarized.

The result is relevant for the application in multi-mode quantum memories, as it allows an assessment of how previously stored quantum states can be moved with high fidelity between register positions, while other quantum states are being read into or out of the superradiant modes. This is a very useful property for recent strategies for quantum computing with qubits encoded in different collective spin modes \cite{tordrup08:holographic,wesenberg09:quantum}. Recent experiments with inhomogeneously broadened spin ensembles and stripline cavities have verified the strong coupling to different collective spin systems \cite{schuster10:high,kubo10:strong}, and we imagine that the gapping and shear effects should be observable in these systems.

\begin{acknowledgments}
K.M. acknowledges support from the EU integrated project AQUTE.
\end{acknowledgments}

\appendix*
\section{Analytical description of the dressed-time states}
\label{sec:analyt-descr-dressed}

In the following, we will establish an analytical expression for the expansion of the dressed-time states (\ref{eq:dresseddef}) on the cavity and bare-time states.

We consider the evolution in terms of the resolvent operator, $\qoP{G}(z) = (z-\qoP{H})^{-1}$ where $z$ is a complex number. By considering the form of $\qoP{G}(z)$ close to the real axis we can compute the forward and backward propagators as
\begin{gather}
  \label{eq:propagators}
  \qoP{G}_{\pm}(w) = \mp i \int e^{i\omega t} \Theta(\pm t) \qoP{U}(t) \,dt
  = \lim_{\eta \to 0^+} \qoP{G}(w \pm i \eta),
\end{gather}
where $\Theta(t)$ is the Heaviside step function \cite{cohen-tannoudji92:atom_photon_inter}. The propagators are in essence Fourier--Laplace transforms of the evolution operator $\qoP{U}$ and are related to the Fourier transform of $\qoP{U}$ by
\begin{gather}
  \label{eq:evolbyresolvent}
  \qoT{U}(\omega) \equiv \int e^{i \omega t} \qoP{U}(t) dt =
  \frac{1}{i} \left[ \qoP{G}_-(\omega) - \qoP{G}_+(\omega) \right].
\end{gather}
In terms of $\qoT{U}$ we can write the dressed-time states, or rather their Fourier transform,
\begin{equation}
  \label{eq:psiwdef}
  \ket{\tilde{\Phi}(\omega)} = \int dt\, e^{i \omega t} \ket{\Phi(t)} =
  \lim_{\tau\to\infty}e^{-i \omega \tau} \qoT{U}(\omega) \ket{-\tau}.
\end{equation}
To calculate a closed expression for $\ket{\tilde{\Phi}(\omega)}$ we partition the reachable part of the single-excitation subspace into the subspace spanned by $\ket{\cavc}$ and that spanned by the bare-time states $\ket{\tau}$, and denote the orthogonal projections onto these subspaces by $\qoP{C}$ and $\qoP{S}$, respectively.

\subsection{Evolution of the cavity state}
\label{sec:evol-cavity-state}

We will first calculate the restriction of $\qoP{G}$ to the subspace spanned by the cavity state $\ket{\cavc}$. Since the subspace is one dimensional, the restriction of any operator is given by a complex number, $\qoP{C} \qoP{G}(z) \qoP{C} = \qoP{C} G_\cavc(z)$ where $G_\cavc(z) = \braket{\cavc|\qoP{G}(z)|\cavc}$.
As $\qoP{G}(z)$ by definition satisfies $(z-\qoP{H}) \qoP{G}(z) = 1$, we have that $\qoP{C} = \qoP{C}(z-\qoP{H})(\qoP{C}+\qoP{S})\qoP{G}(z)\qoP{C}$ and $0=\qoP{S}(z-\qoP{H})(\qoP{C}+\qoP{S}) \qoP{G}(z)\qoP{C}$. Noting that $\qoP{C} H_0 \qoP{S} =0$, it follows that
\begin{align}
  \label{eq:gcavity}
  G_\cavc(z)  &= \Braket{\cavc|
    \frac{\qoP{C}}{z - \qoP{C} \qoP{H}_0 \qoP{C} - \qoP{C} \qoP{R}(z) \qoP{C}}
    |\cavc}\notag\\
  &= \left(z-\omega_\cavc-R_\cavc(z) \right)^{-1}
\end{align}
where $\qoP{R}(z)$ is the level-shift operator,
\begin{equation}
  \label{eq:levelshiftdef}
  \qoP{R}(z) \equiv \qoP{V} + \qoP{V} \frac{\qoP{S}}{\qoP{S} (z-\qoP{H}) \qoP{S}} \qoP{V},
\end{equation}
and $R_\cavc(z) \equiv \braket{\cavc|\qoP{R}(z)|\cavc}$. The level shift can be evaluated explicitly in the basis $\qop{a}_j^\dagger \ket{\text{vac}}$,
\begin{equation}
  \label{eq:rcval}
  R_\cavc(z)  = \Omega^2 \sum_{j=1}^N \frac{\lvert \alpha_j \rvert^2}{z-\omega_j}
  = \Omega^2 \int \frac{\rhoeff(\omega)}{z-\omega} d\omega,
\end{equation}
and it is an analytic function of the complex variable $z$ in the whole complex plane except for the support of $\rhoeff(\omega)$ on the real axis, where $R_\cavc(z)$ has a branch cut such that $R_\cavc(z)$ does not tend to same value from below and from above the cut,
\begin{equation}
  \label{eq:rcpm}
  R_{\cavc \pm}(\omega)
= \lim_{\eta \to 0^+} R_\cavc(\omega \pm i \eta)
= \Omega^2 \left[ \delta_\cavc(\omega) \mp i \gamma_\cavc(\omega)/2\right],
\end{equation}
where $\delta_\cavc$ and $\gamma_\cavc$ are given by Eq.~(\ref{eq:deltacdef}).
Numerically, $R_{\cavc\pm}$ is most easily calculated as $-2\pi i \, \Omega^2\, \mathcal{F}_t\{ \Theta(\pm t) \mathcal{F}^{-1}\{ \rho_{\text{eff}}\}(t) \}$, where $\mathcal F$ denotes the Fourier transform.
By Eq.~(\ref{eq:gcavity}) we can  write $G_{\cavc\pm}(\omega)$ explicitly as
\begin{equation}
  \label{eq:gcpmfinal}
  G_{\cavc\pm}(\omega) = \frac{1}{
   \left[ \omega-\omega_\cavc-\Omega^2 \delta_\cavc(\omega)\right] \pm i\, \Omega^2 \gamma_\cavc(\omega)/2
  },
\end{equation}
from which the exact form of $\braket{\cavc|\qoT{U}(\omega)|\cavc}$ follows from Eq.~(\ref{eq:evolbyresolvent}). We note that $G_{\cavc +}$ is the susceptibility of the cavity mode, and that  $\Delta_\cavc(\omega) = \Omega^2 \delta_\cavc(\omega)$ and $\Gamma_\cavc(\omega)=\Omega^2 \gamma_\cavc(\omega)$ describe the frequency shift of the cavity and the decay from the cavity into the spin ensemble as discussed in greater detail in Refs.~\cite{tsyplyatyev09:dynamics,diniz11:strongly,kurucz10:theory}.

\subsection{Expansion of the dressed-time states}
\label{sec:evol-spin-ensemble}

By algebraic manipulations similar to those used to arrive at Eq.~(\ref{eq:gcavity}), we find that $\qoP{G}_S(z)\equiv\qoP{S} \qoP{G}(z) \qoP{S}$ is related to $G_\cavc$ by
\begin{equation}
  \label{eq:sgs}
  \qoP{G}_S(z)
  = \qoP{G}\topp{0}_S(z) +  \Omega^2 G_\cavc(z) \qoP{G}\topp{0}_S(z) \ket{0}\bra{0}  \qoP{G}\topp{0}_S(z),
\end{equation}
where $\qoP{G}\topp{0}_S(z)$ is the resolvent corresponding to the restriction of $\qoP{H}$ to the spin  subspace, $\qoP{G}\topp{0}_S(z) \equiv \qoP{S}/(z-\qoP{S} \qoP{H} \qoP{S})$ \cite{cohen-tannoudji92:atom_photon_inter}.
Expressing the propagators in terms of the Greens functions as $\qoP{G}\topp{0}_{S\pm}(\omega) = \mp i \int e^{i \omega t} \Theta(\pm t) \qoP{U}\topp{0}_S(t) dt$, we see that
\begin{multline}
  \label{eq:gosmatrix}
  e^{-i \omega \tau} \Braket{0|\qoP{G}\topp{0}_{S\pm}(\omega)|-\tau} \\
  =  \pm \frac{1}{i} \int e^{i \omega (t-\tau)} \Theta(\pm t ) \braket{\tau|t} dt
  \to\begin{cases}
    -i \gamma_\cavc(\omega)\\
    0&
  \end{cases}
\end{multline}
in the limit of $\tau \to \infty$, since we are assuming that the overlap $\braket{0|\tau}$ vanishes for large $\tau$. In other words, the backward propagator does not contribute to the interaction term.
Since $\qoP{S} \ket{\tilde{\Phi}(\omega)} = \lim_{\tau\to\infty}i e^{-i \omega \tau} [{\qoP G_{S+}(\omega)-\qoP G_{S-}(\omega)}] \ket{-\tau}$, it follows from Eqs.~(\ref{eq:sgs}), and (\ref{eq:gosmatrix}), that
\begin{align}
  \label{eq:ketpsires}
  \qoP{S} \ket{\tilde{\Phi}(\omega)}
  &= \int e^{i \omega \tau} \left[
    1 - i\Omega^2 \Theta(\tau) G_{c+}(\omega) \gamma_\cavc(\omega)
  \right] \ket{\tau} d\tau.
\end{align}

To calculate the cavity component of the dressed-time states we note that  $\qoP{C}\qoP{H}_0\qoP{S}=0$, so that it follows from the Dyson expansion $\qoP G(z) = \qoP G^{(0)}(z) + \qoP G(z) \qoP V \qoP G^{(0)}(z)$ that $\qoP{C}\qoP{G}_\pm (\omega)\qoP{S} = \qoP{G}_{C\pm}(\omega)\qoP{V}\qoP{G}\topp{0}_{S\pm}(\omega)$. Then
\begin{multline}
 e^{-i \omega \tau} \qoP{C}\qoP{G}_\pm(\omega)\qoP{S} \ket{-\tau} \\
 = \ket{\cavc} \Omega G_{\cavc\pm}(\omega) e^{-i \omega \tau} \braket{0|  \qoP{G}_{S\pm}\topp{0}(\omega)|-\tau}.
\end{multline}
The matrix element is given by Eq.~(\ref{eq:gosmatrix}), so that
\begin{equation}
 \label{eq:ketpsic}
 \Braket{\cavc|\tilde{\Phi}(\omega)}
 = \Omega G_{c+}(\omega) \gamma_\cavc(\omega).
\end{equation}

Finally, to arrive at Eq.~(\ref{eq:fulldressed}) we note that
\begin{equation}
    1 - i\Omega^2 G_{c+}(\omega) \gamma_\cavc(\omega)
    = \frac{G_{\cavc+}(\omega)}{G_{\cavc-}(\omega)},
\end{equation}
which, since $G_{\cavc-}(\omega)=G_{\cavc+}(\omega)^*$, we can write as $e^{-i\phi(\omega)}$ where $\phi(\omega)=-2\arg[G_{\cavc +}(\omega)]$.


\begin{thebibliography}{10}%
\makeatletter
\providecommand \@ifxundefined [1]{%
 \ifx #1\undefined \expandafter \@firstoftwo
 \else \expandafter \@secondoftwo
\fi
}%
\providecommand \@ifnum [1]{%
 \ifnum #1\expandafter \@firstoftwo
 \else \expandafter \@secondoftwo
\fi
}%
\providecommand \enquote [1]{``#1''}%
\providecommand \bibnamefont  [1]{#1}%
\providecommand \bibfnamefont [1]{#1}%
\providecommand \citenamefont [1]{#1}%
\providecommand\href[0]{\@sanitize\@href}%
\providecommand\@href[1]{\endgroup\@@startlink{#1}\endgroup\@@href}%
\providecommand\@@href[1]{#1\@@endlink}%
\providecommand \@sanitize [0]{\begingroup\catcode`\&12\catcode`\#12\relax}%
\@ifxundefined \pdfoutput {\@firstoftwo}{%
 \@ifnum{\z@=\pdfoutput}{\@firstoftwo}{\@secondoftwo}%
}{%
 \providecommand\@@startlink[1]{\leavevmode}%
 \providecommand\@@endlink[0]{}%
}{%
 \providecommand\@@startlink[1]{%
  \leavevmode
  \pdfstartlink
   attr{/Border[0 0 1 ]/H/I/C[0 1 1]}%
   user{/Subtype/Link/A<</Type/Action/S/URI/URI(#1)>>}%
  \relax
 }%
 \providecommand\@@endlink[0]{\pdfendlink}%
}%
\providecommand \url  [0]{\begingroup\@sanitize \@url }%
\providecommand \@url [1]{\endgroup\@href {#1}{\urlprefix}}%
\providecommand \urlprefix [0]{URL }%
\providecommand \Eprint[0]{\href }%
\@ifxundefined \urlstyle {%
  \providecommand \doi [1]{doi:\discretionary{}{}{}#1}%
}{%
  \providecommand \doi [0]{doi:\discretionary{}{}{}\begingroup
  \urlstyle{rm}\Url }%
}%
\providecommand \doibase [0]{http://dx.doi.org/}%
\providecommand \Doi[1]{\href{\doibase#1}}%
\providecommand \bibAnnote [3]{%
  \BibitemShut{#1}%
  \begin{quotation}\noindent
    \textsc{Key:}\ #2\\\textsc{Annotation:}\ #3%
  \end{quotation}%
}%
\providecommand \bibAnnoteFile [2]{%
  \IfFileExists{#2}{\bibAnnote {#1} {#2} {\input{#2}}}{}%
}%
\providecommand \typeout [0]{\immediate \write \m@ne }%
\providecommand \selectlanguage [0]{\@gobble}%
\providecommand \bibinfo [0]{\@secondoftwo}%
\providecommand \bibfield [0]{\@secondoftwo}%
\providecommand \translation [1]{[#1]}%
\providecommand \BibitemOpen[0]{}%
\providecommand \bibitemStop [0]{}%
\providecommand \bibitemNoStop [0]{.\EOS\space}%
\providecommand \EOS [0]{\spacefactor3000\relax}%
\providecommand \BibitemShut [1]{\csname bibitem#1\endcsname}%
\bibitem{caldeira83:quantum}%
  \BibitemOpen
  \bibfield{author}{%
  \bibinfo {author} {\bibfnamefont{A.~O.}\ \bibnamefont{Caldeira}}\ and\
  \bibinfo {author} {\bibfnamefont{A.~J.}\ \bibnamefont{Leggett}},\ }%
  \bibfield{journal}{%
  \Doi{10.1016/0003-4916(83)90202-6}{\bibinfo {journal} {Ann. Phys.}}\ }%
  \textbf{\bibinfo {volume} {149}},\ \bibinfo {pages} {374} (\bibinfo {year}
  {1983})%
  \bibAnnoteFile{NoStop}{caldeira83:quantum}%
\bibitem{breuer04:non-markovian}%
  \BibitemOpen
  \bibfield{author}{%
  \bibinfo {author} {\bibfnamefont{H.-P.}\ \bibnamefont{Breuer}}, \bibinfo
  {author} {\bibfnamefont{D.}~\bibnamefont{Burgarth}},\ and\ \bibinfo {author}
  {\bibfnamefont{F.}~\bibnamefont{Petruccione}},\ }%
  \bibfield{journal}{%
  \Doi{10.1103/PhysRevB.70.045323}{\bibinfo {journal} {Phys. Rev. B}}\ }%
  \textbf{\bibinfo {volume} {70}},\ \bibinfo {pages} {045323} (\bibinfo {year}
  {2004})%
  \bibAnnoteFile{NoStop}{breuer04:non-markovian}%
\bibitem{schlosshauer08:decoherence}%
  \BibitemOpen
  \bibfield{author}{%
  \bibinfo {author} {\bibfnamefont{M.}~\bibnamefont{Schlosshauer}}, \bibinfo
  {author} {\bibfnamefont{A.~P.}\ \bibnamefont{Hines}},\ and\ \bibinfo {author}
  {\bibfnamefont{G.~J.}\ \bibnamefont{Milburn}},\ }%
  \bibfield{journal}{%
  \Doi{10.1103/PhysRevA.77.022111}{\bibinfo {journal} {Phys. Rev. A}}\ }%
  \textbf{\bibinfo {volume} {77}},\ \bibinfo {pages} {022111} (\bibinfo {year}
  {2008})%
  \bibAnnoteFile{NoStop}{schlosshauer08:decoherence}%
\bibitem{lee54:some}%
  \BibitemOpen
  \bibfield{author}{%
  \bibinfo {author} {\bibfnamefont{T.~D.}\ \bibnamefont{Lee}},\ }%
  \bibfield{journal}{%
  \Doi{10.1103/PhysRev.95.1329}{\bibinfo {journal} {Phys. Rev.}}\ }%
  \textbf{\bibinfo {volume} {95}},\ \bibinfo {pages} {1329} (\bibinfo {year}
  {1954})%
  \bibAnnoteFile{NoStop}{lee54:some}%
\bibitem{scully66:quantum}%
  \BibitemOpen
  \bibfield{author}{%
  \bibinfo {author} {\bibfnamefont{M.}~\bibnamefont{Scully}}\ and\ \bibinfo
  {author} {\bibfnamefont{W.~E.}\ \bibnamefont{Lamb}},\ }%
  \bibfield{journal}{%
  \Doi{PhysRevLett.16.853}{\bibinfo {journal} {Phys. Rev. Lett.}}\ }%
  \textbf{\bibinfo {volume} {16}},\ \bibinfo {pages} {853} (\bibinfo {year}
  {1966})%
  \bibAnnoteFile{NoStop}{scully66:quantum}%
\bibitem{scully67:quantum}%
  \BibitemOpen
  \bibfield{author}{%
  \bibinfo {author} {\bibfnamefont{M.~O.}\ \bibnamefont{Scully}}\ and\ \bibinfo
  {author} {\bibfnamefont{W.~E.}\ \bibnamefont{Lamb}},\ }%
  \bibfield{journal}{%
  \Doi{10.1103/PhysRev.159.208}{\bibinfo {journal} {Phys. Rev.}}\ }%
  \textbf{\bibinfo {volume} {159}},\ \bibinfo {pages} {208} (\bibinfo {year}
  {1967})%
  \bibAnnoteFile{NoStop}{scully67:quantum}%
\bibitem{stenholm00:how}%
  \BibitemOpen
  \bibfield{author}{%
  \bibinfo {author} {\bibfnamefont{S.}~\bibnamefont{Stenholm}},\ }%
  \bibfield{journal}{%
  \Doi{10.1016/S0030-4018(99)00543-X}{\bibinfo {journal} {Opt. Commun.}}\ }%
  \textbf{\bibinfo {volume} {179}},\ \bibinfo {pages} {247} (\bibinfo {year}
  {2000})%
  \bibAnnoteFile{NoStop}{stenholm00:how}%
\bibitem{mollow69:power}%
  \BibitemOpen
  \bibfield{author}{%
  \bibinfo {author} {\bibfnamefont{B.~R.}\ \bibnamefont{Mollow}},\ }%
  \bibfield{journal}{%
  \Doi{10.1103/PhysRev.188.1969}{\bibinfo {journal} {Phys. Rev.}}\ }%
  \textbf{\bibinfo {volume} {188}},\ \bibinfo {pages} {1969} (\bibinfo {year}
  {1969})%
  \bibAnnoteFile{NoStop}{mollow69:power}%
\bibitem{dutt07:quantum}%
  \BibitemOpen
  \bibfield{author}{%
  \bibinfo {author} {\bibfnamefont{M.~V.~G.}\ \bibnamefont{Dutt}}, \bibinfo
  {author} {\bibfnamefont{L.}~\bibnamefont{Childress}}, \bibinfo {author}
  {\bibfnamefont{L.}~\bibnamefont{Jiang}}, \bibinfo {author}
  {\bibfnamefont{E.}~\bibnamefont{Togan}}, \bibinfo {author}
  {\bibfnamefont{J.}~\bibnamefont{Maze}}, \bibinfo {author}
  {\bibfnamefont{F.}~\bibnamefont{Jelezko}}, \bibinfo {author}
  {\bibfnamefont{A.~S.}\ \bibnamefont{Zibrov}}, \bibinfo {author}
  {\bibfnamefont{P.~R.}\ \bibnamefont{Hemmer}},\ and\ \bibinfo {author}
  {\bibfnamefont{M.~D.}\ \bibnamefont{Lukin}},\ }%
  \bibfield{journal}{%
  \Doi{10.1126/science.1139831}{\bibinfo {journal} {Science}}\ }%
  \textbf{\bibinfo {volume} {316}},\ \bibinfo {pages} {1312} (\bibinfo {year}
  {2007})%
  \bibAnnoteFile{NoStop}{dutt07:quantum}%
\bibitem{kurucz09:qubit}%
  \BibitemOpen
  \bibfield{author}{%
  \bibinfo {author} {\bibfnamefont{Z.}~\bibnamefont{Kurucz}}, \bibinfo {author}
  {\bibfnamefont{M.~W.}\ \bibnamefont{S{\o}rensen}}, \bibinfo {author}
  {\bibfnamefont{J.~M.}\ \bibnamefont{Taylor}}, \bibinfo {author}
  {\bibfnamefont{M.~D.}\ \bibnamefont{Lukin}},\ and\ \bibinfo {author}
  {\bibfnamefont{M.}~\bibnamefont{Fleischhauer}},\ }%
  \bibfield{journal}{%
  \Doi{10.1103/PhysRevLett.103.010502}{\bibinfo {journal} {Phys. Rev. Lett.}}\
  }%
  \textbf{\bibinfo {volume} {103}},\ \bibinfo {pages} {010502} (\bibinfo {year}
  {2009})%
  \bibAnnoteFile{NoStop}{kurucz09:qubit}%
\bibitem{petrosyan08:quantum}%
  \BibitemOpen
  \bibfield{author}{%
  \bibinfo {author} {\bibfnamefont{D.}~\bibnamefont{Petrosyan}}\ and\ \bibinfo
  {author} {\bibfnamefont{M.}~\bibnamefont{Fleischhauer}},\ }%
  \bibfield{journal}{%
  \Doi{10.1103/PhysRevLett.100.170501}{\bibinfo {journal} {Phys. Rev. Lett.}}\
  }%
  \textbf{\bibinfo {volume} {100}},\ \bibinfo {pages} {170501} (\bibinfo {year}
  {2008})%
  \bibAnnoteFile{NoStop}{petrosyan08:quantum}%
\bibitem{petrosyan09:reversible}%
  \BibitemOpen
  \bibfield{author}{%
  \bibinfo {author} {\bibfnamefont{D.}~\bibnamefont{Petrosyan}}, \bibinfo
  {author} {\bibfnamefont{G.}~\bibnamefont{Bensky}}, \bibinfo {author}
  {\bibfnamefont{G.}~\bibnamefont{Kurizki}}, \bibinfo {author}
  {\bibfnamefont{I.}~\bibnamefont{Mazets}}, \bibinfo {author}
  {\bibfnamefont{J.}~\bibnamefont{Majer}},\ and\ \bibinfo {author}
  {\bibfnamefont{J.}~\bibnamefont{Schmiedmayer}},\ }%
  \bibfield{journal}{%
  \Doi{10.1103/PhysRevA.79.040304}{\bibinfo {journal} {Phys. Rev. A}}\ }%
  \textbf{\bibinfo {volume} {79}},\ \bibinfo {pages} {040304} (\bibinfo {year}
  {2009})%
  \bibAnnoteFile{NoStop}{petrosyan09:reversible}%
\bibitem{rabl07:molecular}%
  \BibitemOpen
  \bibfield{author}{%
  \bibinfo {author} {\bibfnamefont{P.}~\bibnamefont{Rabl}}\ and\ \bibinfo
  {author} {\bibfnamefont{P.}~\bibnamefont{Zoller}},\ }%
  \bibfield{journal}{%
  \Doi{10.1103/PhysRevA.76.042308}{\bibinfo {journal} {Phys. Rev. A}}\ }%
  \textbf{\bibinfo {volume} {76}},\ \bibinfo {pages} {042308} (\bibinfo {year}
  {2007})%
  \bibAnnoteFile{NoStop}{rabl07:molecular}%
\bibitem{tordrup08:holographic}%
  \BibitemOpen
  \bibfield{author}{%
  \bibinfo {author} {\bibfnamefont{K.}~\bibnamefont{Tordrup}}, \bibinfo
  {author} {\bibfnamefont{A.}~\bibnamefont{Negretti}},\ and\ \bibinfo {author}
  {\bibfnamefont{K.}~\bibnamefont{M{\o}lmer}},\ }%
  \bibfield{journal}{%
  \Doi{10.1103/PhysRevLett.101.040501}{\bibinfo {journal} {Phys. Rev. Lett.}}\
  }%
  \textbf{\bibinfo {volume} {101}},\ \bibinfo {pages} {040501} (\bibinfo {year}
  {2008})%
  \bibAnnoteFile{NoStop}{tordrup08:holographic}%
\bibitem{wesenberg09:quantum}%
  \BibitemOpen
  \bibfield{author}{%
  \bibinfo {author} {\bibfnamefont{J.~H.}\ \bibnamefont{Wesenberg}}, \bibinfo
  {author} {\bibfnamefont{A.}~\bibnamefont{Ardavan}}, \bibinfo {author}
  {\bibfnamefont{G.~A.~D.}\ \bibnamefont{Briggs}}, \bibinfo {author}
  {\bibfnamefont{J.~J.~L.}\ \bibnamefont{Morton}}, \bibinfo {author}
  {\bibfnamefont{R.~J.}\ \bibnamefont{Schoelkopf}}, \bibinfo {author}
  {\bibfnamefont{D.~I.}\ \bibnamefont{Schuster}},\ and\ \bibinfo {author}
  {\bibfnamefont{K.}~\bibnamefont{M{\o}lmer}},\ }%
  \bibfield{journal}{%
  \Doi{10.1103/PhysRevLett.103.070502}{\bibinfo {journal} {Phys. Rev. Lett.}}\
  }%
  \textbf{\bibinfo {volume} {103}},\ \bibinfo {pages} {070502} (\bibinfo {year}
  {2009})%
  \bibAnnoteFile{NoStop}{wesenberg09:quantum}%
\bibitem{imamoglu08:cavity-qed}%
  \BibitemOpen
  \bibfield{author}{%
  \bibinfo {author} {\bibfnamefont{A.}~\bibnamefont{Imamoglu}},\ }%
  \bibfield{journal}{%
  \Doi{10.1103/PhysRevLett.102.083602}{\bibinfo {journal} {Phys. Rev. Lett.}}\
  }%
  \textbf{\bibinfo {volume} {102}},\ \bibinfo {pages} {083602} (\bibinfo {year}
  {2009})%
  \bibAnnoteFile{NoStop}{imamoglu08:cavity-qed}%
\bibitem{herskind09:realization}%
  \BibitemOpen
  \bibfield{author}{%
  \bibinfo {author} {\bibfnamefont{P.~F.}\ \bibnamefont{Herskind}}, \bibinfo
  {author} {\bibfnamefont{A.}~\bibnamefont{Dantan}}, \bibinfo {author}
  {\bibfnamefont{J.~P.}\ \bibnamefont{Marler}}, \bibinfo {author}
  {\bibfnamefont{M.}~\bibnamefont{Albert}},\ and\ \bibinfo {author}
  {\bibfnamefont{M.}~\bibnamefont{Drewsen}},\ }%
  \bibfield{journal}{%
  \Doi{10.1038/nphys1302}{\bibinfo {journal} {Nat. Phys.}}\ }%
  \textbf{\bibinfo {volume} {5}},\ \bibinfo {pages} {494} (\bibinfo {year}
  {2009})%
  \bibAnnoteFile{NoStop}{herskind09:realization}%
\bibitem{anderson55:spin}%
  \BibitemOpen
  \bibfield{author}{%
  \bibinfo {author} {\bibfnamefont{A.~G.}\ \bibnamefont{Anderson}}, \bibinfo
  {author} {\bibfnamefont{R.~L.}\ \bibnamefont{Garwin}}, \bibinfo {author}
  {\bibfnamefont{E.~L.}\ \bibnamefont{Hahn}}, \bibinfo {author}
  {\bibfnamefont{J.~W.}\ \bibnamefont{Horton}}, \bibinfo {author}
  {\bibfnamefont{G.~L.}\ \bibnamefont{Tucker}},\ and\ \bibinfo {author}
  {\bibfnamefont{R.~M.}\ \bibnamefont{Walker}},\ }%
  \bibfield{journal}{%
  \Doi{10.1063/1.1721903}{\bibinfo {journal} {J. Appl. Phys.}}\ }%
  \textbf{\bibinfo {volume} {26}},\ \bibinfo {pages} {1324} (\bibinfo {year}
  {1955})%
  \bibAnnoteFile{NoStop}{anderson55:spin}%
\bibitem{carlson83:storage}%
  \BibitemOpen
  \bibfield{author}{%
  \bibinfo {author} {\bibfnamefont{N.~W.}\ \bibnamefont{Carlson}}, \bibinfo
  {author} {\bibfnamefont{L.~J.}\ \bibnamefont{Rothberg}}, \bibinfo {author}
  {\bibfnamefont{A.~G.}\ \bibnamefont{Yodh}}, \bibinfo {author}
  {\bibfnamefont{W.~R.}\ \bibnamefont{Babbitt}},\ and\ \bibinfo {author}
  {\bibfnamefont{T.~W.}\ \bibnamefont{Mossberg}},\ }%
  \bibfield{journal}{%
  \Doi{10.1364/OL.8.000483}{\bibinfo {journal} {Opt. Lett.}}\ }%
  \textbf{\bibinfo {volume} {8}},\ \bibinfo {pages} {483} (\bibinfo {year}
  {1983})%
  \bibAnnoteFile{NoStop}{carlson83:storage}%
\bibitem{kraus06:quantum}%
  \BibitemOpen
  \bibfield{author}{%
  \bibinfo {author} {\bibfnamefont{B.}~\bibnamefont{Kraus}}, \bibinfo {author}
  {\bibfnamefont{W.}~\bibnamefont{Tittel}}, \bibinfo {author}
  {\bibfnamefont{N.}~\bibnamefont{Gisin}}, \bibinfo {author}
  {\bibfnamefont{M.}~\bibnamefont{Nilsson}}, \bibinfo {author}
  {\bibfnamefont{S.}~\bibnamefont{Kroll}},\ and\ \bibinfo {author}
  {\bibfnamefont{J.~I.}\ \bibnamefont{Cirac}},\ }%
  \bibfield{journal}{%
  \Doi{10.1103/PhysRevA.73.020302}{\bibinfo {journal} {Phys. Rev. A}}\ }%
  \textbf{\bibinfo {volume} {73}},\ \bibinfo {pages} {020302(R)} (\bibinfo
  {year} {2006})%
  \bibAnnoteFile{NoStop}{kraus06:quantum}%
\bibitem{afzelius09:multimode}%
  \BibitemOpen
  \bibfield{author}{%
  \bibinfo {author} {\bibfnamefont{M.}~\bibnamefont{Afzelius}}, \bibinfo
  {author} {\bibfnamefont{C.}~\bibnamefont{Simon}}, \bibinfo {author}
  {\bibfnamefont{H.}~\bibnamefont{de~Riedmatten}},\ and\ \bibinfo {author}
  {\bibfnamefont{N.}~\bibnamefont{Gisin}},\ }%
  \bibfield{journal}{%
  \Doi{10.1103/PhysRevA.79.052329}{\bibinfo {journal} {Phys. Rev. A}}\ }%
  \textbf{\bibinfo {volume} {79}},\ \bibinfo {pages} {052329} (\bibinfo {year}
  {2009})%
  \bibAnnoteFile{NoStop}{afzelius09:multimode}%
\bibitem{wu09:storage}%
  \BibitemOpen
  \bibfield{author}{%
  \bibinfo {author} {\bibfnamefont{H.}~\bibnamefont{Wu}}, \bibinfo {author}
  {\bibfnamefont{R.~E.}\ \bibnamefont{George}}, \bibinfo {author}
  {\bibfnamefont{J.~H.}\ \bibnamefont{Wesenberg}}, \bibinfo {author}
  {\bibfnamefont{K.}~\bibnamefont{M{\o}lmer}}, \bibinfo {author}
  {\bibfnamefont{D.~I.}\ \bibnamefont{Schuster}}, \bibinfo {author}
  {\bibfnamefont{R.~J.}\ \bibnamefont{Schoelkopf}}, \bibinfo {author}
  {\bibfnamefont{K.~M.}\ \bibnamefont{Itoh}}, \bibinfo {author}
  {\bibfnamefont{A.}~\bibnamefont{Ardavan}}, \bibinfo {author}
  {\bibfnamefont{J.~J.~L.}\ \bibnamefont{Morton}},\ and\ \bibinfo {author}
  {\bibfnamefont{G.~A.~D.}\ \bibnamefont{Briggs}},\ }%
  \bibfield{journal}{%
  \Doi{10.1103/PhysRevLett.105.140503}{\bibinfo {journal} {Phys. Rev. Lett.}}\
  }%
  \textbf{\bibinfo {volume} {105}},\ \bibinfo {pages} {140503} (\bibinfo {year}
  {2010})%
  \bibAnnoteFile{NoStop}{wu09:storage}%
\bibitem{holstein40:field}%
  \BibitemOpen
  \bibfield{author}{%
  \bibinfo {author} {\bibfnamefont{T.}~\bibnamefont{Holstein}}\ and\ \bibinfo
  {author} {\bibfnamefont{H.}~\bibnamefont{Primakoff}},\ }%
  \bibfield{journal}{%
  \Doi{10.1103/PhysRev.58.1098}{\bibinfo {journal} {Phys. Rev.}}\ }%
  \textbf{\bibinfo {volume} {58}},\ \bibinfo {pages} {1098} (\bibinfo {year}
  {1940})%
  \bibAnnoteFile{NoStop}{holstein40:field}%
\bibitem{bloch30:ferromagnetismus}%
  \BibitemOpen
  \bibfield{author}{%
  \bibinfo {author} {\bibfnamefont{F.}~\bibnamefont{Bloch}},\ }%
  \bibfield{journal}{%
  \Doi{10.1007/BF01339661}{\bibinfo {journal} {Z. Phys.}}\ }%
  \textbf{\bibinfo {volume} {61}},\ \bibinfo {pages} {206} (\bibinfo {year}
  {1930})%
  \bibAnnoteFile{NoStop}{bloch30:ferromagnetismus}%
\bibitem{schuster10:high}%
  \BibitemOpen
  \bibfield{author}{%
  \bibinfo {author} {\bibfnamefont{D.~I.}\ \bibnamefont{Schuster}}, \bibinfo
  {author} {\bibfnamefont{A.~P.}\ \bibnamefont{Sears}}, \bibinfo {author}
  {\bibfnamefont{E.}~\bibnamefont{Ginossar}}, \bibinfo {author}
  {\bibfnamefont{L.}~\bibnamefont{DiCarlo}}, \bibinfo {author}
  {\bibfnamefont{L.}~\bibnamefont{Frunzio}}, \bibinfo {author}
  {\bibfnamefont{J.~J.~L.}\ \bibnamefont{Morton}}, \bibinfo {author}
  {\bibfnamefont{H.}~\bibnamefont{Wu}}, \bibinfo {author}
  {\bibfnamefont{G.~A.~D.}\ \bibnamefont{Briggs}}, \bibinfo {author}
  {\bibfnamefont{B.~B.}\ \bibnamefont{Buckley}}, \bibinfo {author}
  {\bibfnamefont{D.~D.}\ \bibnamefont{Awschalom}},\ and\ \bibinfo {author}
  {\bibfnamefont{R.~J.}\ \bibnamefont{Schoelkopf}},\ }%
  \bibfield{journal}{%
  \Doi{10.1103/PhysRevLett.105.140501}{\bibinfo {journal} {Phys. Rev. Lett.}}\
  }%
  \textbf{\bibinfo {volume} {105}},\ \bibinfo {pages} {140501} (\bibinfo {year}
  {2010})%
  \bibAnnoteFile{NoStop}{schuster10:high}%
\bibitem{kubo10:strong}%
  \BibitemOpen
  \bibfield{author}{%
  \bibinfo {author} {\bibfnamefont{Y.}~\bibnamefont{Kubo}}, \bibinfo {author}
  {\bibfnamefont{F.~R.}\ \bibnamefont{Ong}}, \bibinfo {author}
  {\bibfnamefont{P.}~\bibnamefont{Bertet}}, \bibinfo {author}
  {\bibfnamefont{D.}~\bibnamefont{Vion}}, \bibinfo {author}
  {\bibfnamefont{V.}~\bibnamefont{Jacques}}, \bibinfo {author}
  {\bibfnamefont{D.}~\bibnamefont{Zheng}}, \bibinfo {author}
  {\bibfnamefont{A.}~\bibnamefont{Dr{\'e}au}}, \bibinfo {author}
  {\bibfnamefont{J.~F.}\ \bibnamefont{Roch}}, \bibinfo {author}
  {\bibfnamefont{A.}~\bibnamefont{Auffeves}}, \bibinfo {author}
  {\bibfnamefont{F.}~\bibnamefont{Jelezko}}, \bibinfo {author}
  {\bibfnamefont{J.}~\bibnamefont{Wrachtrup}}, \bibinfo {author}
  {\bibfnamefont{M.~F.}\ \bibnamefont{Barthe}}, \bibinfo {author}
  {\bibfnamefont{P.}~\bibnamefont{Bergonzo}},\ and\ \bibinfo {author}
  {\bibfnamefont{D.}~\bibnamefont{Esteve}},\ }%
  \bibfield{journal}{%
  \Doi{10.1103/PhysRevLett.105.140502}{\bibinfo {journal} {Phys. Rev. Lett.}}\
  }%
  \textbf{\bibinfo {volume} {105}},\ \bibinfo {pages} {140502} (\bibinfo {year}
  {2010})%
  \bibAnnoteFile{NoStop}{kubo10:strong}%
\bibitem{cohen-tannoudji92:atom_photon_inter}%
  \BibitemOpen
  \bibfield{author}{%
  \bibinfo {author} {\bibfnamefont{C.}~\bibnamefont{Cohen-Tannoudji}}, \bibinfo
  {author} {\bibfnamefont{J.}~\bibnamefont{Dupont-Roc}},\ and\ \bibinfo
  {author} {\bibfnamefont{G.}~\bibnamefont{Grynberg}},\ }%
  \emph{\bibinfo {title} {Atom-Photon Interactions}}\ (\bibinfo {publisher}
  {John Wiley and Sons},\ \bibinfo {year} {1992})%
  \bibAnnoteFile{NoStop}{cohen-tannoudji92:atom_photon_inter}%
\bibitem{tsyplyatyev09:dynamics}%
  \BibitemOpen
  \bibfield{author}{%
  \bibinfo {author} {\bibfnamefont{O.}~\bibnamefont{Tsyplyatyev}}\ and\
  \bibinfo {author} {\bibfnamefont{D.}~\bibnamefont{Loss}},\ }%
  \bibfield{journal}{%
  \Doi{10.1103/PhysRevA.80.023803}{\bibinfo {journal} {Phys. Rev. A}}\ }%
  \textbf{\bibinfo {volume} {80}},\ \bibinfo {pages} {023803} (\bibinfo {year}
  {2009})%
  \bibAnnoteFile{NoStop}{tsyplyatyev09:dynamics}%
\bibitem{diniz11:strongly}%
  \BibitemOpen
  \bibfield{author}{%
  \bibinfo {author} {\bibfnamefont{I.}~\bibnamefont{Diniz}}, \bibinfo {author}
  {\bibfnamefont{S.}~\bibnamefont{Portolan}}, \bibinfo {author}
  {\bibfnamefont{R.}~\bibnamefont{Ferreira}}, \bibinfo {author}
  {\bibfnamefont{J.~M.}\ \bibnamefont{G{\'e}rard}}, \bibinfo {author}
  {\bibfnamefont{P.}~\bibnamefont{Bertet}},\ and\ \bibinfo {author}
  {\bibfnamefont{A.}~\bibnamefont{Auff{\`e}ves}},\ }%
  \enquote{\bibinfo {title} {Strongly coupling a cavity to inhomogeneous
  ensembles of emitters : potential for long lived solid-state quantum
  memories},}\  (\bibinfo {year} {2011}),\
  \Eprint{http://arxiv.org/abs/1101.1842}{arxiv:1101.1842}%
  \bibAnnoteFile{NoStop}{diniz11:strongly}%
\bibitem{kurucz10:theory}%
  \BibitemOpen
  \bibfield{author}{%
  \bibinfo {author} {\bibfnamefont{Z.}~\bibnamefont{Kurucz}}, \bibinfo {author}
  {\bibfnamefont{J.~H.}\ \bibnamefont{Wesenberg}},\ and\ \bibinfo {author}
  {\bibfnamefont{K.}~\bibnamefont{M{\o}lmer}},\ }%
  \enquote{\bibinfo {title} {Spectroscopic properties of inhomogeneously
  broadened spin ensembles in a cavity},}\  (\bibinfo {year} {2011}),\
  \Eprint{http://arxiv.org/abs/1101.4828}{arxiv:1101.4828}%
  \bibAnnoteFile{NoStop}{kurucz10:theory}%
\end{thebibliography}
\end{document}